\documentclass[aps,prl,twocolumn,superscriptaddress]{revtex4}
\usepackage[pdftex]{graphicx}

\begin{document}

\title{Traveling wave deceleration of heavy polar molecules in low-field seeking states}

\author{N.E. Bulleid}
\author{R.J. Hendricks}
\author{E.A. Hinds}
\affiliation{Centre for Cold Matter, Blackett Laboratory, Imperial College London, Prince Consort Road, London, SW7 2AZ. United Kingdom.}
\author{Samuel A. Meek}
\author{Gerard Meijer}
\affiliation{Fritz-Haber-Institut der Max-Planck-Gesellschaft, Faradayweg 4-6, 14195 Berlin, Germany.}
\author{Andreas Osterwalder}
\affiliation{Ecole Polytechnique F\'ed\'erale de Lausanne, Institut des Sciences et Ing\'enierie Chimiques, 1015 Lausanne, Switzerland.}
\author{M.R. Tarbutt}
\affiliation{Centre for Cold Matter, Blackett Laboratory, Imperial College London, Prince Consort Road, London, SW7 2AZ. United Kingdom.}




\begin{abstract}
We demonstrate the deceleration of heavy polar molecules in low-field seeking states by combining a cryogenic source and a travelling-wave Stark decelerator. The cryogenic source provides a high intensity beam with low speed and temperature, and the travelling-wave decelerator provides large deceleration forces and high phase-space acceptance. We prove these techniques using YbF molecules and find the experimental data to be in excellent agreement with numerical simulations. These methods extend the scope of Stark deceleration to a very wide range of molecules.
\end{abstract}


\maketitle

The ability to control the speed of a molecular beam has been valuable for a wide range of applications. This new control arose from the development of Stark deceleration \cite{Bethlem(1)99} in which a time-varying inhomogeneous electric field is used to deliver molecules with a precisely-tuned velocity in the range 0-1000~m/s, and with a spread of velocities as low as 1~m/s. These velocity-controlled beams have been used for high-resolution spectroscopy \cite{vanVeldhoven(1)04}, measurements of collision cross-sections with exceptional energy resolution \cite{Gilijamse(1)06, Scharfenberg(1)11}, and for precise tests of fundamental physics \cite{Hudson(1)06}. Once decelerated to rest, molecules can be stored in electric \cite{Bethlem(1)00}, magnetic \cite{Sawyer(1)07}, or ac traps \cite{vanVeldhoven(1)05}, where the lifetimes of long-lived states can be measured \cite{Meerakker(1)05, Gilijamse(1)07}, and the collision physics of the trapped molecules can be studied \cite{Sawyer(1)08, Parazzoli11}. Magnetic \cite{Vanhaecke(1)07, Narevicius(1)08} and optical \cite{Fulton(1)06} analogues of the Stark decelerator have also been developed.

\begin{figure}
\begin{centering}
\includegraphics[width=0.45\textwidth]{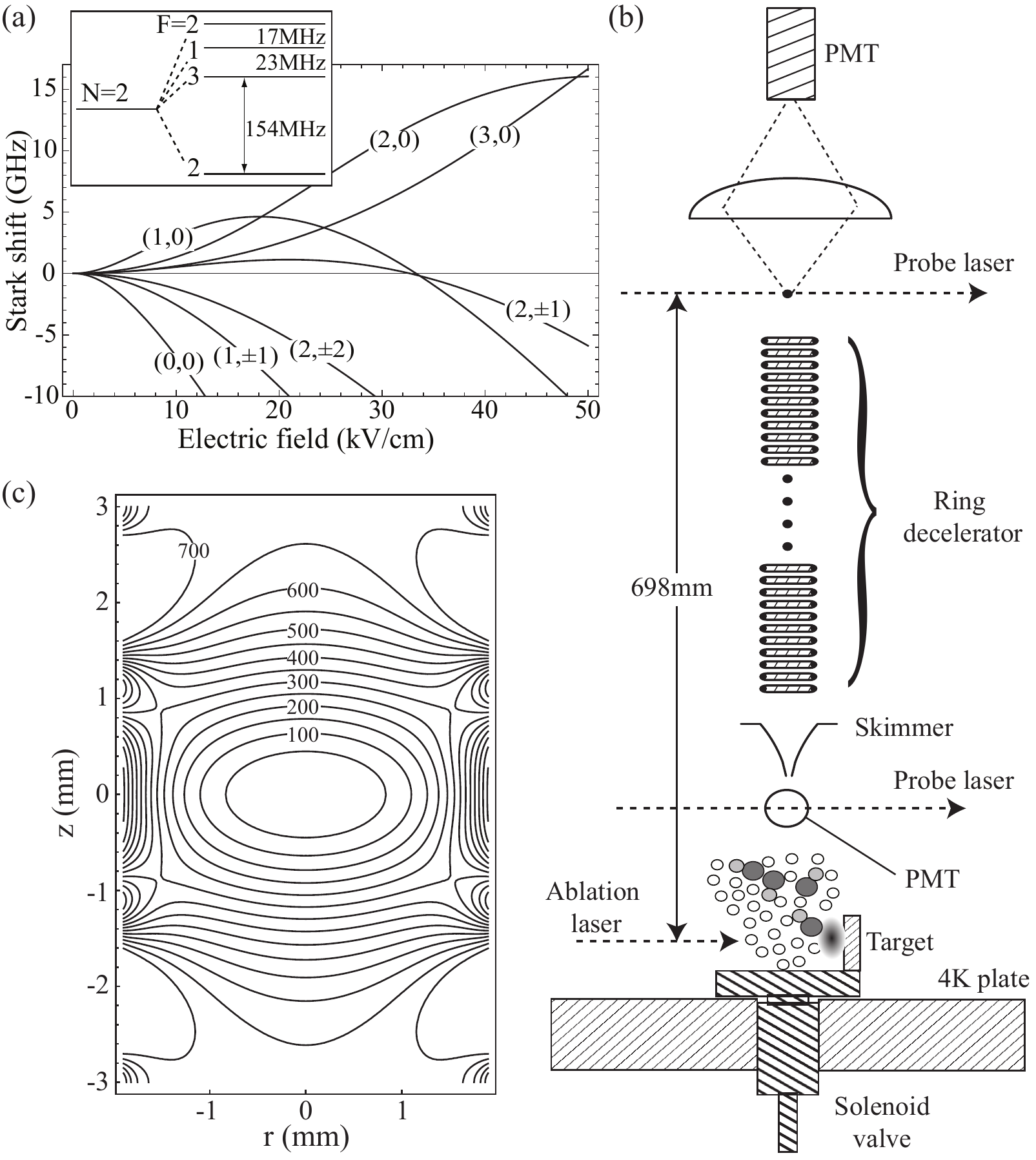}
\caption{\label{fig:Experimentalsetup}(a) Stark shifts of low-lying rotational states $(N,M_{N})$ of YbF, and hyperfine components of $N=2$ (inset). (b) Schematic of the experiment, not to scale. (c) Contour plot of the potential energy of YbF $(N=2,M_{N}=0)$ molecules in a plane through the centre of the ring decelerator. Contours are drawn at 50\,mK intervals, and labelling is in mK. The traps move along the decelerator at a speed governed by the frequency of the applied sinusoidal voltages.}
\end{centering}
\end{figure}

There is a strong desire to extend deceleration techniques to heavier molecules. These are becoming increasingly important in tests of fundamental physics, such as the measurement of the electron's electric dipole moment \cite{Hudson(1)11} and tests of parity violation in nuclei \cite{DeMille(1)08} and chiral molecules \cite{Darquie(1)10}. These measurements all rely on heavy molecules, and their sensitivities could be greatly increased by using slower and colder beams \cite{Tarbutt(1)09}. Also important is control over the motion of biomolecules, which will enhance spectroscopic studies of these basic building blocks of life, and allow their separation according to rotational state and conformational structure. Deceleration of these heavy molecules in a conventional Stark decelerator poses several difficulties. First, the kinetic energy to be removed is large and so a very large number of deceleration stages is needed. Second, transverse focussing of the molecules becomes ineffective at low speed resulting in a severe loss of molecules from the decelerator, particularly for the long decelerators needed for heavier molecules \cite{Sawyer(2)08}. Third, the molecules need to be in a low-field seeking state so that they are focussed by the decelerator in both longitudinal and transverse directions \cite{Bethlem(2)00, Meerakker(1)06}, but heavy molecules are only low-field seeking when the field is small. This is illustrated in Fig.\,\ref{fig:Experimentalsetup}(a) which shows the Stark shifts of the low-lying energy levels of YbF. Those states that are low-field seeking become high-field seeking as the field increases, and this severely limits the amount of energy that can be removed in each deceleration stage. Deceleration in high-field seeking states has been demonstrated using the alternating gradient focussing method \cite{Tarbutt(1)04, Wohlfart(1)08}, but has not been pursued because the phase space acceptance is low and the construction and operation are particularly difficult \cite{Tarbutt(1)08, Tarbutt(1)09}.

Recent developments make it possible to solve all these problems. A new type of Stark decelerator has been developed where low-field seeking molecules are captured in traveling, gradually decelerating, three-dimensional traps \cite{Osterwalder(1)10, Meek(1)11}. The molecules are confined in the same trap throughout, so there is no loss. The trapping fields are relatively low, and can be kept below the turn-over point of the selected low-field seeking state, but the decelerating force and phase-space acceptance are large. Concurrently, new molecular beam sources have been developed using cryogenic buffer gas methods \cite{Maxwell(1)05, Patterson(1)07}. They produce intense beams of cold molecules with low velocities, thus reducing the kinetic energy to be removed by deceleration. Here, we report the traveling-wave deceleration of YbF molecules produced in a cryogenic source. The methods we demonstrate are applicable to a wide range of heavy molecules, but deceleration of YbF is particularly relevant as these molecules have recently been used to measure the electron's electric dipole moment \cite{Hudson(1)11}.

Figure \ref{fig:Experimentalsetup}(b) shows the experimental setup. Helium gas expands supersonically from a pulsed solenoid valve, modified for low temperature use and connected to the 4\,K cold head of a closed-cycle cryocooler. Laser ablation of a target composed of AlF$_3$ and Yb produces YbF which is entrained in the helium flow. The molecular pulses have a duration of about $60$~$\mu$s, a translational temperature of 6\,K, and a mean speed of 315\,m/s. This speed is considerably higher than the 204\,m/s terminal speed expected for a 4\,K helium expansion, and we attribute this to local heating of the solenoid valve by a few Kelvin. Far slower beams can be obtained from an effusive or partially hydrodynamic buffer gas cell, but they produce much longer pulses that are less well suited to the small size scale of the decelerator. Charcoal sorption pumps, also cooled to 4\,K, provide high pumping speed in the source region, allowing a high throughput of helium and a correspondingly high molecular beam intensity. The molecules pass through the 3\,mm opening of a skimmer 143\,mm above the valve nozzle, and enter the decelerator whose first ring is centred 172\,mm above the nozzle. The decelerator \cite{Osterwalder(1)10, Meek(1)11} is 480\,mm long and is built from a series of 4\,mm diameter ring electrodes, spaced by 1.5\,mm. A sinusoidal time-varying potential is applied to each ring, with an amplitude of 10\,kV and a 45$^{\circ}$ phase shift between adjacent rings. This produces a series of potential energy minima that travel along the decelerator at a speed governed by the applied frequency. Figure \ref{fig:Experimentalsetup}(c) shows one potential minimum for YbF in the $(N,M_{N})=(2,0)$ state. Molecules are loaded into a traveling trap by turning the potentials on when they reach the position of the first minimum. Lowering the applied frequency decelerates the trap and the molecules within. The maximum electric field in the trap is $37$\,kV/cm and the trap depth is maximized by using molecules in the $N=2$ rotational state which has the largest Stark shift at this field, as shown in Fig.\,\ref{fig:Experimentalsetup}(a). After leaving the decelerator, the molecules in this state are detected by laser-induced fluorescence (LIF) following excitation on the $A^2\Pi_{\frac{1}{2}}(v=0) \leftarrow X^2\Sigma^+(v=0)$ Q(2) transition at 552\,nm. The $N=2$ level has four hyperfine components, as shown in Fig.\,\ref{fig:Experimentalsetup}(a), which are partly resolved in the spectrum, and we tune the probe laser to detect molecules in the upper two components ($F=1,2$). Of the 8 $M_{F}$ sub-components of these two states, half correlate to $M_{N}=0$ at high field and the other half to $M_{N}=\pm 1$. Only those in $M_{N}=0$ are strongly focussed through the decelerator. A second LIF detector between source and skimmer is used to measure the source flux.

To help interpret our results, we simulate molecular trajectories through the apparatus. In these simulations we fix the longitudinal distribution of molecules in the source, which is the only uncertain parameter in the experiment, to be a Gaussian with a full width at half maximum of 9.4\,mm. All other parameters are set to the values they are known to have in the experiments. Molecules in all the $M_{N}$-substates of $N=2$ are simulated, and we find that about 90\% of the signal comes from those in $M_{N}=0$ and the rest from those in $M_{N}=\pm 1$.

\begin{figure}[t]
\begin{centering}
\includegraphics[width=0.45\textwidth]{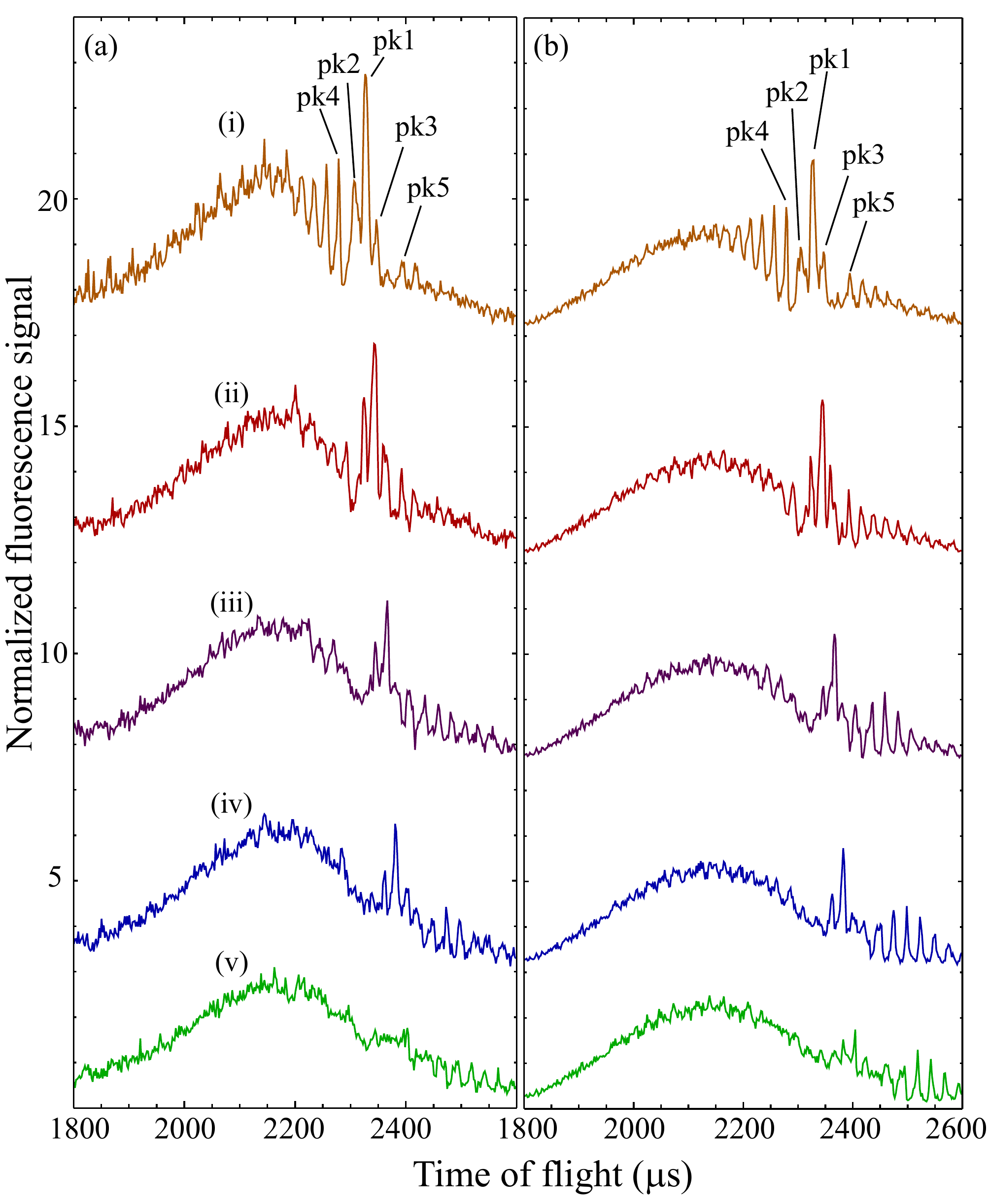}
\caption{\label{fig:N=2deceleration} Time-of-flight distributions of YbF molecules in $N=2$, (a) measured and (b) simulated. Profiles are offset vertically for clarity. The decelerator is set to decelerate molecules from an initial speed of 300~m\,s$^{-1}$. Decelerations and final speeds of the decelerated molecules are (i) 0~m\,s$^{-2}$, 300~m\,s$^{-1}$, (ii) 3710~m\,s$^{-2}$, 294~m\,s$^{-1}$, (iii) 7350~m\,s$^{-2}$, 288~m\,s$^{-1}$, (iv) 10900~m\,s$^{-2}$, 282~m\,s$^{-1}$, (v) 14400~m\,s$^{-2}$, 276~m\,s$^{-1}$.}
\end{centering}
\end{figure}

Figure~\ref{fig:N=2deceleration} shows the measured time-of-flight profiles for various applied decelerations, along with simulation results. The waveforms are constructed to trap molecules with an initial speed of 300\,m/s and decelerate them by a chosen amount. For the experiments and simulations, the vertical scale is the signal obtained with the decelerator on divided by the amplitude of the profile obtained with the decelerator voltage turned off, with no additional scaling. All the features observed in the experiments are faithfully reproduced by the simulations, showing the high degree of control and understanding obtained. There is some difference between experiment and simulation regarding the size of the signal, most noticeable in profile (i), which we attribute to the fact that the normalization data were taken later and so were reduced in amplitude because of a slow decline in signal due to target degradation.

\begin{figure}[!t]
\begin{centering}
\includegraphics[width=0.45\textwidth]{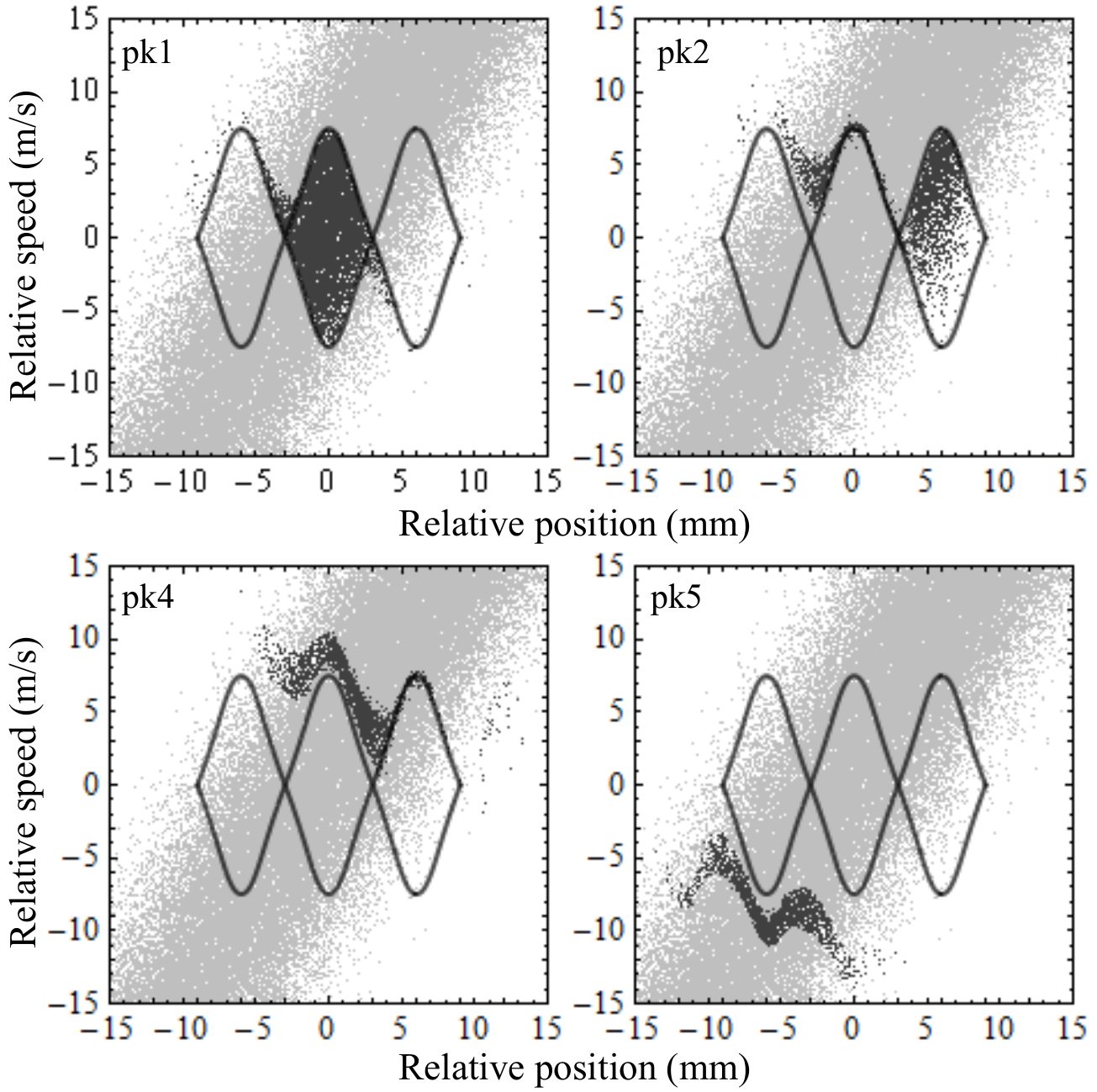}
\caption{\label{fig:phasespaceplots} Light points: longitudinal positions and speeds of simulated molecules at the moment when the decelerator is turned on. Dark points: those molecules that produce the labelled peaks in Fig.\,\ref{fig:N=2deceleration}. Bold lines: region that encloses the trapped molecules (the separatrix), calculated analytically.}
\end{centering}
\end{figure}

The top trace shows the signal obtained with a frequency of 25\,kHz applied to guide molecules at a constant speed of 300\,m/s. There is a background of molecules whose arrival times are unaffected by the decelerator and produce the same Gaussian profile as when the decelerator is off. Although they are not trapped in the decelerator, they are radially confined, and so we observe more molecules when the decelerator is on. The large narrow peak in the profile (labelled pk1) is due to molecules loaded into the trap and transported through the decelerator as intended. The two smaller peaks on either side (pk2 and pk3) are separated from the main peak by $\pm 20$\,$\mu$s, corresponding to a distance of $\pm 6$\,mm, and are due to molecules loaded into traps that are ahead and behind the main one. We see this clearly in Fig.~\ref{fig:phasespaceplots} which shows the longitudinal positions and speeds, relative to the central trap, of all the simulated molecules at the moment when the decelerator is turned on. The bold lines in the plots enclose the regions in phase space where molecules are trapped, and are calculated analytically. The molecules initially occupy a diagonal band in this space, with the faster ones at the front (i.e. on the right) and the distribution spanning about three traps. The dark points show which molecules form each of the labelled peaks in Fig.~\ref{fig:N=2deceleration}. We see that pk1 comes from molecules in the central trap while pk2 comes from those in the trap ahead. Though not shown in Fig.~\ref{fig:phasespaceplots}, pk3 is clearly from molecules in the trap behind the central one. Because the central trap is the most densely filled, pk1 is the largest peak. The peak labelled pk4 is due to molecules that are slightly too fast to be trapped, but that lie very close to the separatrix, as shown in Fig.~\ref{fig:phasespaceplots}. Their velocity relative to the traps is strongly modulated by the periodic potential and they spend a long time near the trap maxima where their relative speed is lowest. When they leave the decelerator these molecules are bunched around the potential maximum that lies 15\,mm ahead of the central trap, and they arrive as a narrow bunch at the detector. This is the first in a series of equally-spaced peaks all of which are due to un-trapped molecules that are bunched up around successive maxima of the potential. The bunching is less significant for higher relative velocities, and so the earlier peaks in the series are smaller. Similarly, pk5 is due to molecules that are too slow to be trapped but lie close to the separatrix and are bunched at the potential maxima. Their speed is well below the central speed of the distribution, and so the peak is small.

The lower traces in Fig.~\ref{fig:N=2deceleration} show that when the applied deceleration is increased the trapped molecules arrive later as they are decelerated from 300~m/s to 276~m/s. There are always three bunches of decelerated molecules, corresponding to the three traps that are filled at the beginning. The decelerated bunches also get smaller because the phase space acceptance decreases with increasing deceleration. The bunches of un-trapped molecules on the left side disappear, while the ones on the right side become more prominent, because the trap velocity is shifting to lower speeds and so the bunching becomes more effective for the slower un-trapped molecules. We repeated these experiments for molecules in $N=3$ and $N=1$. The results for $N=3$ were similar to those in Fig.\,\ref{fig:N=2deceleration}, but the signals and the maximum obtainable deceleration were both smaller due to the smaller Stark shift (see Fig.\,\ref{fig:Experimentalsetup}(a)). For $N=1$ the signal was very small and only the trapped molecules were transmitted, the un-trapped ones being ejected because they are high-field seekers in the high-field regions of the decelerator.

\begin{figure}[!t]
\begin{centering}
\includegraphics[width=0.45\textwidth]{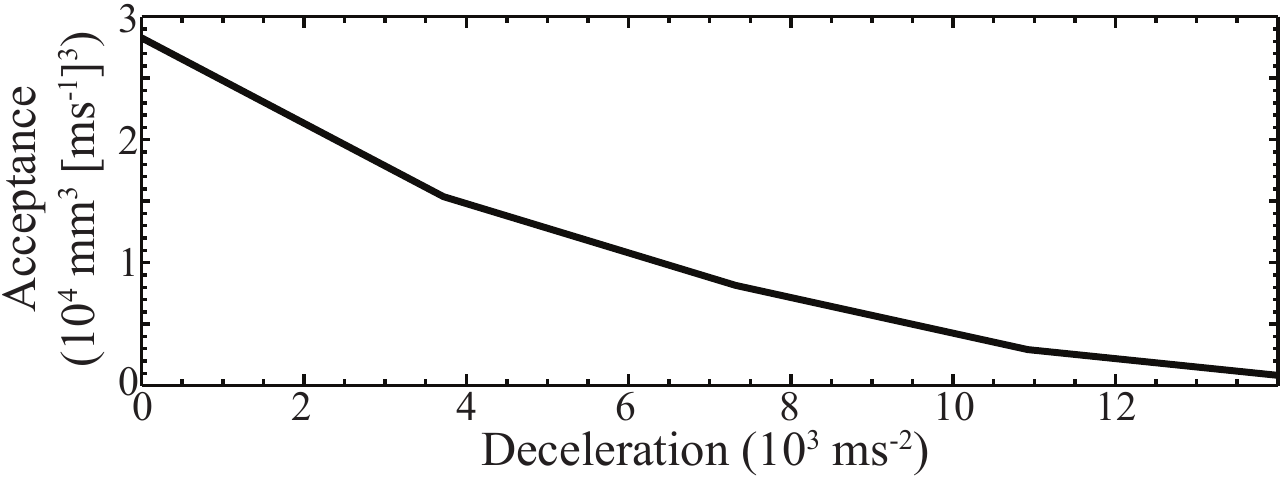}
\caption{\label{fig:acceptance}Phase-space acceptance of the decelerator versus deceleration for YbF in $(N,M_{N})=(2,0)$.}
\end{centering}
\end{figure}

Figure \ref{fig:acceptance} shows the calculated phase-space acceptance of the decelerator for YbF molecules in the $(2,0)$ state. We simulate trajectories through a long decelerator for 10\,ms, using a uniform initial distribution that completely fills the acceptance volume. The fraction of molecules in the decelerated bunch measures the acceptance. It is 28000\,mm$^{3}$\,[m\,s$^{-1}$]$^{3}$ when guiding the molecules, and falls to 4200\,mm$^{3}$\,[m\,s$^{-1}$]$^{3}$ when the deceleration is $10^{4}$\,m\,s$^{-2}$. The deceleration of YbF was considered in \cite{Tarbutt(1)09} where two decelerators based on previous designs were analyzed in detail. The present decelerator offers many advantages over these other decelerators. The maximum electric field is 6 times smaller, the construction is simpler, and the acceptance is more than 10 times larger for the same deceleration. Since the molecules are trapped in the moving potential well from the outset, there are no additional losses at low speed, and no losses associated with coupling from the decelerator into a trap. The novel cryogenic source used here provides short, intense, pulses of cold YbF at relatively low speed, and is well-suited for use with the decelerator. From this source, the phase-space density of YbF molecules in the $(2,0)$ state is estimated to be 4\,mm$^{-3}$\,[m\,s$^{-1}$]$^{-3}$. In recent work we have increased this by a factor of 5, and with further development we expect that the initial speed can be reduced to 200\,m\,s$^{-1}$ or lower, without compromising on the other important parameters. A 2\,m long decelerator, operated at $10^{4}$\,m\,s$^{-2}$ would then bring about $10^{5}$ molecules to rest. Following deceleration to low speed, a short period of laser cooling \cite{Shuman(1)10} could be applied to reduce the remaining velocity spread in all directions \cite{Wall(1)11}. YbF is a good example of a molecule that is amenable to laser cooling \cite{Zhuang(1)11}, with only a few laser wavelengths required. Since the velocity spread exiting the decelerator is only a few m\,s$^{-1}$, a thousand scattered photons is sufficient to cool the molecules well below 1\,mK. This combination of intense cryogenic sources, travelling wave deceleration, and laser cooling, is likely to be a powerful method for delivering heavy polar molecules at ultracold temperatures for a wide range of future experiments.

\begin{acknowledgments}
We thank Ben Sauer and Danny Segal for valuable discussions and Jon Dyne and Steve Maine for technical support. This work was supported by the UK EPSRC and the Royal Society and has received funding from the European Community's Seventh Framework Programme FP7/2007-2013 under grant 216774 and ERC-2009-AdG under grant 247142-MolChip.
\end{acknowledgments}

\end{document}